\documentclass[prl,twocolumn,superscriptaddress]{revtex4-1}%
\usepackage[ansinew]{inputenc}
\usepackage{graphicx}
\usepackage{amsmath}
\usepackage{amsthm}
\usepackage{bm}
\usepackage{layout}
\usepackage{float}
\usepackage{amsfonts}
\usepackage{amssymb}
\usepackage{txfonts}%
\begin{document}

\title{Are there fundamental limits for observing quantum phenomena from within
quantum theory?}

\begin{abstract}
Does there exist a limit for the applicability of quantum theory for objects
of large mass or size, or objects whose states are of large complexity or
dimension of the Hilbert space? The possible answers range from practical
limitations due to decoherence within quantum theory to fundamental limits due
to collapse models that modify quantum theory. Here, we suggest the viewpoint
that there might be also fundamental limits \textit{without altering the
quantum laws}. We first demonstrate that for two quantum spins systems of a
given spin length, no violation of local realism can be observed, if the
measurements are sufficiently coarse-grained. Then we show that there exists a
fundamental limit for the precision of measurements due to (i) the Heisenberg
uncertainty relation which has to be applied to the measuring apparatus, (ii)
relativistic causality, and (iii) the finiteness of resources in any
laboratory including the whole universe. This suggests that there might exist
a limit for the size of the systems (dimension of the Hilbert space) above
which no violation of local realism can be seen anymore.

\end{abstract}

%

\author{Johannes Kofler}%
%

\affiliation
{Institut f\"ur Quantenoptik und Quanteninformation, \"Osterreichische Akademie der Wissenschaften,\\ Boltzmanngasse 3, 1090 Wien, Austria}%
%

\author{{\v C}aslav Brukner}%
%

\affiliation{Fakultät für Physik, Universit\"{a}%
t Wien, Boltzmanngasse 5, 1090 Wien, Austria}
%

\maketitle

Despite the enormous success of quantum physics and its wide range of
applications, the region of the whole parameter space over which the validity
of quantum physics has been directly tested is still rather modest. In
Ref.~\cite{Legg2002}, Leggett argues that, taking for example the length
scale, it is commonly claimed that quantum laws are valid down to the Planck
scale ($\sim\!10^{-35}\,$m) and up to the size of the characteristic length
scale of the Universe ($\sim\!10^{+27}\,$m). This results in 62 orders of
magnitude, compared to about 25 orders of magnitude over which the theory has
been directly tested so far. Notwithstanding recent experimental
achievements~\cite{Arnd1999,Frie2000,Buls2001} that could demonstrate quantum
interference in large systems, it remains an open question: \textit{Are there
principal limitations on observing quantum phenomena of objects of large mass
or size, or objects whose states are of large complexity or dimension of the
Hilbert space?}

Here we suggest a possible affirmative answer to the above question when
considering the dimensionality of the Hilbert space. This explanation differs
conceptually from decoherence~\cite{Zure1991,Zure2003} or collapse
theories~\cite{Ghir1986,Penr1986}. Fully within quantum theory, our approach
puts the emphasis on the \textit{observability} of quantum effects and shows
that the necessary measurement accuracy to see such effects in systems of
sufficiently large Hilbert space dimension cannot be met because of the
\textit{conjunction of quantum physics itself, relativity theory, and the
finiteness of resources in any laboratory}.

To illustrate our idea, let us start by investigating the experimental
requirements for achieving a violation of local realism for systems of
increasing Hilbert space dimension. Such a violation in a Bell
experiment~\cite{Bell1964} is generally accepted as a genuine quantum
phenomenon. We consider 2 spin-$s$ particles in a generalized singlet state%
\begin{equation}
\left\vert \psi\right\rangle =\frac{1}{\sqrt{2s+1}}\,%
{\displaystyle\sum\limits_{m=-s}^{s}}
(-1)^{s-m}\left\vert m\right\rangle _{A}\left\vert -m\right\rangle _{B},
\label{general singlet}%
\end{equation}
where, $\left\vert m\right\rangle _{A}$\ ($\left\vert m\right\rangle _{B}$)
denotes the eigenstates of the spin operator's $z$-component of Alice (Bob).
Measuring spin components on either side, this state allows to violate local
realism for arbitrarily large $s$ (dimension $2s+1$) but it is necessary that
the \textit{inaccuracy of the angle settings} of Alice and Bob, $\Delta\theta
$, is at most in the order of the inverse spin size: $\Delta\theta
\lesssim\frac{1}{s}$. This is the case in the Clauser-Horne-Shimony-Holt-type
inequality~\cite{Clau1969} used in Ref.~\cite{Pere1995}, where the difference
between setting angles has to be about $\frac{1}{2s+1}$ as well as in
Ref.~\cite{Merm1980} where the setting angle has to fulfil $0<\sin
\theta\approx\theta<\frac{1}{2s}$.

Meanwhile more efficient inequalities for higher-dimensional systems have been
found~\cite{Coll2002,Jung2009}. So one could argue that for the
state~(\ref{general singlet}) there might exist Bell inequalities which do not
require such a strict condition as $\Delta\theta\lesssim\frac{1}{s}$. However,
it has been shown by Peres that for a resolution which is much worse than the
intrinsic quantum uncertainty of a spin coherent state, i.e.
\begin{equation}
\Delta\theta\gg\frac{1}{\sqrt{s}},
\end{equation}
all Bell inequalities will necessarily be satisfied for the
state~(\ref{general singlet}), since the correlations between outcomes of
inaccurate measurements become \textit{(classical) correlations between
classical spins}~\cite{Pere1995}. This approach was extended to the time
evolution of quantum systems and the concept of \textit{macroscopic realism}
as introduced by Leggett and Garg \cite{Legg1985}. It was shown that for
\textquotedblleft classical Hamiltonians\textquotedblright\ and under the
restriction of coarse-grained measurements, an arbitrarily large quantum spin
evolves as an ensemble of classical spins following a classical mechanical
evolution~\cite{Kofl2007}.

We will now extend the result of Peres to \textit{arbitrary states} of two
spin-$s$ systems, taking the restriction of coarse-grained measurements where
\textit{neighboring spin directions} cannot be distinguished. We first
introduce the basic mathematical concepts for the further analysis. The
(normalized and positive) $Q$-distribution~\cite{Agar1981} of a two-system
state $\hat{\rho}_{AB}$ is given by%
\begin{equation}
Q_{AB}(\Omega_{A},\Omega_{B})\equiv\left(  \dfrac{2s+1}{4\pi}\right)
^{\!2}\,\langle\Omega_{A},\Omega_{B}|\hat{\rho}_{AB}|\Omega_{A},\Omega
_{B}\rangle
\end{equation}
with $\Omega_{i}$ the spin direction and $|\Omega_{i}\rangle$ the spin
coherent~\cite{Radc1971} states for system $i=A,B$. In a coarse-grained spin
measurement of system $i$, the whole unit sphere is decomposed into a number
of mutually disjoint angular regions (\textquotedblleft
slots\textquotedblright) $\Omega_{i}^{(k)}$, labeled by $k$. (The
decompositions for $A$ and $B$ need not be the same.) A positive operator
valued measurement (POVM) on system $i$ has the elements~\cite{Kofl2008}%
\begin{equation}
\hat{P}_{i}^{(k)}\equiv\dfrac{2s+1}{4\pi}\,%
{\displaystyle\iint\nolimits_{\Omega_{i}^{(k)}}}
|\Omega_{i}\rangle\langle\Omega_{i}|\,\text{d}^{2}\Omega_{i}%
\end{equation}
which correspond to these coarse-grained slots ($%
{\textstyle\sum\nolimits_{k}}
\hat{P}_{i}^{(k)}=\openone$). The joint probability to find the outcome $m$
for system $A$ and the outcome $n$ for system $B$ is given by $w_{AB}%
^{(mn)}=\,$Tr$[\hat{\rho}_{AB}\hat{P}_{A}^{(m)}\hat{P}_{B}^{(n)}]$ or,
equivalently, just via integration over the (positive and normalized)
$Q$-distribution:
\begin{equation}
w_{AB}^{(mn)}=%
{\displaystyle\iint\nolimits_{\Omega_{A}^{(m)}}}
{\displaystyle\iint\nolimits_{\Omega_{B}^{(n)}}}
Q_{AB}(\Omega_{A},\Omega_{B})\,\text{d}^{2}\Omega_{A}\,\text{d}^{2}\Omega_{B}.
\end{equation}
(Please note that in general $Q_{AB}(\Omega_{A},\Omega_{B})$ does not
factorize, i.e.\ it cannot be written as a product $Q_{A}(\Omega_{A}%
)Q_{B}(\Omega_{B})$ of two $Q$-functions of the individual systems.) Upon
measurement, the state $\hat{\rho}_{AB}$ is reduced to $\hat{\rho}_{AB}%
^{(mn)}=\hat{M}_{A}^{(m)}\hat{M}_{B}^{(n)}\hat{\rho}_{AB}\hat{M}_{A}^{(m)\dag
}\hat{M}_{B}^{(n)\dag}/w_{AB}^{(mn)}$, with $\hat{M}_{i}^{(k)}$ the Kraus
operators obeying $\hat{M}_{i}^{(k)\dag}\hat{M}_{i}^{(k)}=\hat{P}_{i}^{(k)}$.
The corresponding $Q$-distribution of the reduced state is $Q_{AB}%
^{(mn)}(\Omega_{A},\Omega_{B})=(\tfrac{2s+1}{4\pi})^{2}\,\langle\Omega
_{A},\Omega_{B}|\hat{\rho}_{AB}^{(mn)}|\Omega_{A},\Omega_{B}\rangle$.

Under the restriction of \textit{sufficiently} coarse-grained measurements
where the (polar and azimuthal) angular size of these regions, $\Delta\Theta$,
has to be much larger than the inverse square root of the spin length $s$,
$\Delta\Theta\gg1/\!\sqrt{s}$, the $Q$-distribution before measurement is very
well approximated by the (weighted) mixture of the $Q$-distributions of the
possible reduced states $\hat{\rho}_{AB}^{(mn)}$~\cite{Kofl2008,foot2}:%
\begin{equation}
Q_{AB}(\Omega_{A},\Omega_{B})\approx%
{\displaystyle\sum\nolimits_{m}}
{\displaystyle\sum\nolimits_{n}}
w_{AB}^{(mn)}\,Q_{AB}^{(mn)}(\Omega_{A},\Omega_{B}).
\end{equation}

Moreover, this condition holds for \textit{all} possible \textquotedblleft
setting choices\textquotedblright\ of decompositions for the angular regions
of the systems $A$ and $B$. We could also decompose the regions into a
different set of mutually disjoint regions, denoted by $\bar{\Omega}%
_{i}^{(k^{\prime})}$ (where the decompositions of $A$ and $B$ need not be the
same). Then we would get the similar condition $Q_{AB}(\Omega_{A},\Omega
_{B})\approx%
{\textstyle\sum\nolimits_{m^{\prime}}}
{\textstyle\sum\nolimits_{n^{\prime}}}
w_{AB}^{(m^{\prime}n^{\prime})}\,\bar{Q}_{AB}^{(m^{\prime}n^{\prime})}%
(\Omega_{A},\Omega_{B})$, where the $\bar{Q}_{AB}^{(m^{\prime}n^{\prime})}$
are the $Q$-functions for the reduced states under decomposition into
$\bar{\Omega}_{i}^{(k^{\prime})}$. This means that under sufficiently
coarse-grained measurements one can consider all results as stemming from an
underlying probability distribution, representing a classical ensemble of
spins~\cite{Kofl2007,Kofl2008}. In particular, there exists a \textit{joint}
(positive and normalized) probability $p^{(mm^{\prime}nn^{\prime})}\equiv
p(\Omega_{A}^{(m)},\bar{\Omega}_{A}^{(m^{\prime})},\Omega_{B}^{(n)}%
,\bar{\Omega}_{B}^{(n^{\prime})})$ for the (potential) values corresponding
simultaneously to $\Omega_{A}^{(m)}$ and $\bar{\Omega}_{A}^{(m^{\prime})}$ for
spin $A$ and $\Omega_{B}^{(n)}$ and $\bar{\Omega}_{B}^{(n^{\prime})}$\ for
spin $B$, which is given by the integration over the intersections of the
corresponding regions:%
\begin{equation}
p^{(mm^{\prime}nn^{\prime})}=%
{\displaystyle\iint\nolimits_{\Omega_{A}^{(m)}\cap\bar{\Omega}_{A}%
^{(m^{\prime})}}}
{\displaystyle\iint\nolimits_{\Omega_{B}^{(n)}\cap\bar{\Omega}_{B}%
^{(m^{\prime})}}}
Q_{AB}(\Omega_{A},\Omega_{B})\,\text{d}^{2}\Omega_{A}\,\text{d}^{2}\Omega_{B}.
\end{equation}

All the above can of course be easily generalized to more than two different
compositions for each system and to more than two systems. $Q_{AB}$ can be
understood as providing a probability distribution over local hidden variables
$(\Omega_{A},\Omega_{B})$. Under sufficiently coarse-grained spin measurements
$\Delta\Theta\gg1/\!\sqrt{s}$, no Bell inequality can be violated, as a joint
probability distribution exists. Therefore, the criterion for having a chance
to see deviations from a fully classical description of the two spins reads%
\begin{equation}
\Delta\theta\lesssim\frac{1}{\sqrt{s}}\,. \label{eq delta theta v2}%
\end{equation}

It is clear that for increasingly large $s$ it is hard to meet this
experimental requirement and violate local realism or see non-classical
correlations. But in fact, we suggest that there might exist even a
fundamental upper limit on $s$---stemming from the Heisenberg uncertainty
principle, relativity theory, and finiteness of resources---up to which, for a
given measurement device, one can still see non-classical correlations.

The measurements are done with Stern-Gerlach magnets or similar devices. The
angle of a magnet has to be set with an accuracy $\Delta\theta$. The
Heisenberg uncertainty implies $\Delta L\,\Delta\theta\geq\hbar/2$, where
$\Delta L$ is the intrinsic uncertainty of the angular momentum of the whole
magnet and $\hbar\sim10^{-34}\,$Js it the reduced Planck constant. Note that
in general the form of the angular momentum uncertainty relation is
state-dependent as $\theta$ is $2\pi$-periodic and its variance is naturally
bounded from above~\cite{Pere1995,Pegg2005}. However, in our case of a well
aligned measurement apparatus, $\theta$ is sharply peaked with a very small
width $\Delta\theta\ll1$ and the problems associated with the periodicity of
the angle variable can be neglected. Therefore, the commutator of the angle
and angular momentum operator can be written as%
\begin{equation}
\lbrack\hat{\theta},\hat{L}]=\text{i}\,\hbar\,. \label{eq comm}%
\end{equation}

In Ref.~\cite{Calm2004} it was shown that the Planck length is a device
independent limit which determines the inaccuracy of any distance measurement.
Following these thoughts, we can derive a bound on the angular inaccuracy
$\Delta\theta$ from within quantum physics. First assume that the spin enters
the inhomogeneous magnetic field of the Stern-Gerlach magnet at time $t=0$ and
leaves the interaction zone at time $\tau$. The Hamiltonian of a
\textit{freely rotating} magnet is%
\begin{equation}
\hat{H}=\frac{\hat{L}^{2}}{2I}\,, \label{eq Hamiltonian}%
\end{equation}
where $\hat{L}$ is the angular momentum operator of the magnet and $I\sim
MR^{2}$ its moment of inertia with $M$ and $R$ the mass and characteristic
size, respectively. (We often neglect factors of the order of 1 throughout our
derivations.) In the Heisenberg picture, the time evolution of the polar angle
is given by the Heisenberg equation of motion d$\hat{\theta}/$d$t=-$%
i$\,[\hat{\theta},\hat{H}]/\hbar=\hat{L}/I$. Therefore, for the measurement
duration $\tau$, $\hat{\theta}(\tau)=\hat{\theta}(0)+\hat{L}\tau/I$, where
$\hat{L}$ is independent of time. We recall the Robertson inequality $\Delta
A\,\Delta B\geq\frac{1}{2}\,|\langle\lbrack\hat{A},\hat{B}]\rangle
|$~\cite{Robe1929}, which holds for any two observables $\hat{A}$ and $\hat
{B}$. Using the commutation relation~(\ref{eq comm}), we obtain%
\begin{equation}
\Delta\theta(0)\,\Delta\theta(\tau)\geq\frac{\hbar\tau}{2I}\sim\dfrac
{\hbar\tau}{MR^{2}}\,. \label{ineq thth}%
\end{equation}
It follows that at least \textit{one of the two} quantities, $\hat{\theta}(0)$
and $\hat{\theta}(\tau)$, has a spread of%
\begin{equation}
\Delta\theta\gtrsim\dfrac{1}{R}\sqrt{\frac{\hbar\tau}{M}}\,, \label{ineq dth}%
\end{equation}
which is denoted as the \textit{standard quantum limit}.

Using condition (\ref{eq delta theta v2}), we obtain the constraint on the
spin size such that non-classicality can possibly be seen:%
\begin{equation}
s\lesssim\dfrac{MR^{2}}{\hbar\tau}\,. \label{ineq s}%
\end{equation}
Choosing typical laboratory values $R\sim1\,$m, $M\sim1\,$kg, $\tau\sim1\,$s,
one arrives at $s\lesssim10^{34}$.

In order to obtain a fundamental limit on $s$, we follow Ref.~\cite{Calm2004}
and impose physical constraints:

By \textit{relativistic causality} the operative size $R$ of the freely moving
measurement device cannot exceed the distance that light can travel during the
interaction time $\tau$: $R\leq c\tau$, with $c\sim10^{8}\,$m/s the speed of
light. Note that this effective measurement apparatus not only contains the
Stern-Gerlach magnet but also the table on which it is mounted and possibly
the whole earth etc. Using this constraint, ineq.~(\ref{ineq dth}) becomes%
\begin{equation}
\Delta\theta\gtrsim\sqrt{\frac{\hbar}{cMR}}\,, \label{ineq dth2}%
\end{equation}
and ineq.~(\ref{eq delta theta v2}) then reads%
\begin{equation}
s\lesssim\dfrac{cMR}{\hbar}\,. \label{ineq s 2}%
\end{equation}
Taking again $R\sim1\,$m, $M\sim1\,$kg, this leads to $s\lesssim10^{42}$.

As a \textit{fundamental limit} one can choose as size and mass of the device
the radius and mass of the observable universe, $R_{U}\sim10^{27}\,$m and
$M_{U}\sim10^{53}\,$kg, respectively. This leads to the condition%
\begin{equation}
s\lesssim\frac{cM_{U}R_{U}}{\hbar}\sim10^{122}\,. \label{ineq univ1}%
\end{equation}

Note that under the stronger accuracy condition $\Delta\theta\lesssim\frac
{1}{s}$ for the state (\ref{general singlet}), the spin size for which the
Bell inequalities of Refs.~\cite{Pere1995,Merm1980} can be violated is only
$s\lesssim10^{61}$. Both limits are exceedingly large. However, insofar as the
size and mass of the universe as ultimate resources are finite, there is a
fundamental limit on how large the Hilbert space of the systems can be such
that one is still able to observes genuine quantum features. Despite the fact
that the Hilbert space for a spin $10^{61}$ ($10^{122}$) can be formed by only
about 200 (400) qubits, the question whether or not these two limits are
trivial depends on whether a \textit{physical} spin (measured by a
Stern-Gerlach apparatus) of such size can in principle be formed.

In order to additionally \textit{avoid gravitational collapse}, the size of
the measurement apparatus must be larger than the Schwarzschild radius
corresponding to its mass $M$: $R\geq2GM/c^{2}$, with $G\sim10^{-10}\,$m$^{3}%
$kg$^{-1}$s$^{-2}$ the gravitational constant~\cite{foot}. Using this
constraint, ineq.~(\ref{ineq dth2}) becomes%
\begin{equation}
\Delta\theta\gtrsim\frac{l_{P}}{R}\,,
\end{equation}
and ineq.~(\ref{eq delta theta v2}) then reads%
\begin{equation}
s\lesssim\dfrac{R^{2}}{l_{P}^{2}}\,, \label{cond 2}%
\end{equation}
where $l_{P}\equiv\sqrt{\hbar G/c^{3}}\sim10^{-35}\,$m is the Planck length.
This limit can intuitively be understood since the inaccuracy in the
measurement of an angle (which is the ratio of two distances) of, say, a rod
of length $R$ is essentially given by the inaccuracy in the position
measurement of its extremal point (given by the Planck length) divided by the
length of the rod. The latter, of course, has an uncertainty $\Delta R$
itself, but this leads to a negligible higher order effect.

For $R\sim1\,$m, we get $s\lesssim10^{70}$. As an alternative
\textit{fundamental limit} one can again take the size of the universe
$R_{U}\sim10^{27}\,$m, which leads to the condition%
\begin{equation}
s\lesssim\dfrac{R_{U}^{2}}{l_{P}^{2}}\sim10^{124}\,. \label{ineq univ2}%
\end{equation}
The limit for $s$ in conditions (\ref{ineq univ1}) and (\ref{ineq univ2})
being similar, reflects the fact that our observable universe is close to be a
black hole.

Several remarks have to be made at this point:

\begin{enumerate}
\item[(i)] The fact that the standard quantum limit can be beaten by
contractive states~\cite{Yuen1983,Cave1985,Giov2004} does not change the
validity of inequality~(\ref{ineq dth}) as \textit{two subsequent}
measurements are still bounded by~(\ref{ineq thth})~\cite{Calm2004}.

\item[(ii)] The assumption of a free time evolution (\ref{eq Hamiltonian})
must be justified. One could imagine a large setup with all kinds of fields
and rods and clever mechanisms which compensates movements of the magnet
within itself. But such a construction terminates at the causal radius.

\item[(iii)] One might argue that only \textit{angle differences} are
important in the Bell experiment and not the local angles themselves. This
fact, however, cannot be exploited as the two (space-time) measurement regions
must be \textit{space-like separated} and no rigid connection can exist.

\item[(iv)] We did not take into account \textit{other inaccuracies}, in
particular the ones in position and momentum of the spin particles, in the
inhomogeneous magnetic field, the state preparation, the ones during the
measurement procedure on the screen after the magnet, and inaccuracies in the
reference frames of Alice and Bob themselves. All these components of the
experimental setup have to obey the Heisenberg uncertainty as well and maybe
impose a much stricter limit. In this sense, we have derived a very
conservative \textit{upper bound} on the maximal spin length ($cMR/\hbar$ or
$R^{2}/l_{P}^{2}$, respectively) beyond which it is impossible to observe the
quantum features of an arbitrary state.

\item[(v)] We note that a violation of Bell's inequality remains possible for
arbitrarily large spin size $s$, if the two parties Alice and Bob can perform
arbitrary unitary transformations before their measurements even if the latter
are still coarse-grained~\cite{Kofl2008,Son2009}. Consider, for example, the
macroscopically entangled state $(\left\vert s\right\rangle _{A}\left\vert
-s\right\rangle _{B}+\left\vert -s\right\rangle _{A}\left\vert s\right\rangle
_{B})/\sqrt{2}$. For observing a violation of Bell's inequality, it is
sufficient that Alice and Bob perform coarse-grained which-hemisphere
measurements on their local spin systems, but then it is necessary that they
have the ability to produce Schrödinger cat-like states of the form
$\left\vert \pm\alpha\right\rangle _{A}=\cos\alpha\left\vert s\right\rangle
_{A}\pm\sin\alpha\left\vert -s\right\rangle _{A}$. Such a combination of a
\textquotedblleft non-classical\textquotedblright\ transformation and a
(\textquotedblleft classical\textquotedblright) coarse-grained measurement is
effectively not a coarse-grained measurement in which \textit{neighboring spin
directions in real configuration space are bunched together}. Such measurement
observables are denoted as \textquotedblleft unreasonable\textquotedblright%
~\cite{Pere1995}. (Reasonable coarse-grained observables correspond to
measurements that bunch together those outcomes that are neighboring in real space.)
\end{enumerate}

\textit{Conclusion}. We demonstrated that a violation of local realism cannot
be seen for spins of a certain size because of the Heisenberg uncertainty
relation, relativistic causality, (gravitational collapse,) and the finiteness
of resources in any laboratory. The view taken by most scientists is that the
concepts of physical theories being established due to and verified by
experiments are independent of the amount of physical resources needed to
carry out these experiments. In stark contrast, Benioff~\cite{Beni2003} and
Davies~\cite{Davi2007} recently argued that physical laws should not be
treated as infinitely precise, immutable mathematical constructs, but must
rather respect the finiteness of resources in the universe. This might impose
a fundamental limit on the precision of the laws and the specifiability of
physical states. We enforce this view by proposing that quantum mechanics
itself puts a limit on the possibility to observe quantum phenomena if only a
restricted amount of physical resources is available.

\textit{Acknowledgments}. We thank M.~Aspelmeyer, T.~Paterek, and A.~Zeilinger
for helpful discussions. This work was supported by the Austrian Science
Foundation FWF within Project No.\ P19570-N16, SFB and CoQuS No.\ W1210-N16.


\begin{thebibliography}{99}
\bibitem {foot}It is of course conceivable that an observer performs
experiments inside a black hole. Such an observer, however, cannot communicate
his results to regions outside the black hole.
\end{thebibliography}
\end{document}